\documentclass{article}%
\usepackage{amsfonts}
\usepackage{amsmath}
\usepackage{amssymb}
\usepackage{graphicx}%
\newtheorem{theorem}{Theorem}

\newtheorem{definition}[theorem]{Definition}

\newtheorem{lemma}[theorem]{Lemma}

\newtheorem{remark}[theorem]{Remark}

\newenvironment{proof}[1][Proof]{\noindent\textbf{#1.} }{\ \rule{0.5em}{0.5em}}
\begin{document}

\title{New applications of pseudoanalytic function theory to the Dirac equation}
\author{Antonio Casta\~{n}eda, Vladislav V. Kravchenko\\Secci\'{o}n de Posgrado e Investigaci\'{o}n\\Escuela Superior de Ingenier\'{\i}a Mec\'{a}nica y El\'{e}ctrica\\Instituto Polit\'{e}cnico Nacional\\C.P.07738 M\'{e}xico D.F., \\MEXICO\\phone: 525557296000, ext. 54782\\e-mail: vkravchenko@ipn.mx}
\maketitle

\begin{abstract}
In the present work we establish a simple relation between the Dirac equation
with a scalar and an electromagnetic potentials in a two-dimensional case and
a pair of decoupled Vekua equations. In general these Vekua equations are
bicomplex. However we show that the whole theory of pseudoanalytic functions
without modifications can be applied to these equations under a certain not
restrictive condition. As an example we formulate the similarity principle
which is the central reason why a pseudoanalytic function and as a consequence
a spinor field depending on two space variables share many of the properties
of analytic functions.

One of the surprising consequences of the established relation with
pseudoanalytic functions consists in the following result. Consider the Dirac
equation with a scalar potential depending on one variable with fixed energy
and mass. In general this equation cannot be solved explicitly even if one
looks for wave functions of one variable. Nevertheless for such Dirac equation
we obtain an algorithmically simple procedure for constructing in explicit
form a complete system of exact solutions (depending on two variables). These
solutions generalize the system of powers $1,z,z^{2},\ldots$ in complex
analysis and are called formal powers. With their aid any regular solution of
the Dirac equation can be represented by its Taylor series in formal powers.

\end{abstract}

\section{Introduction}

The Dirac equation with a fixed energy and the Vekua equation describing
pseudoanalytic (=generalized analytic) functions both are first order elliptic
systems, and it would be quite natural to expect a deep interrelation between
their theories especially in the case when all potentials and wave functions
in the Dirac equation depend on two space variables only. Nevertheless there
is not much work done in this direction$\footnote{We refer to the work
\cite{Battle} where the theory of pseudoanalytic functions was used in a way
completely different from ours for studying the two-dimensional Dirac equation
with a scalar or a pseudoscalar potential.}$ due to the fact that any of the
traditional matrix representations of the Dirac operator does not allow to
visualize a relation between the Dirac equation in the two-dimensional case
and the Vekua equation. The Dirac equation is a system of four complex
equations which does not decouple in a two-dimensional situation but decouples
in the one-dimensional case only.

In the present work we establish a simple relation between the Dirac equation
with a scalar and an electromagnetic potentials in a two-dimensional case from
one side and a pair of decoupled Vekua equations from the other. As a first
step we use the matrix transformation proposed in \cite{Krbag} (see also
\cite{AQA} and \cite{KSbook}) which allows us to rewrite the Dirac equation in
a covariant form as a biquaternionic equation. This is not our aim to discuss
here the advantages of our biquaternionic reformulation of the Dirac equation
compared with other its representations (the interested reader can find some
of the arguments in \cite{AQA}). We point out only that our transformation is
$\mathbb{C}$-linear as well as the resulting Dirac operator, which is not the
case for a better known biquaternionic reformulation of the Dirac operator
introduced by C. Lanczos in \cite{Lanczos} (see \cite{Gsponer} and \cite{AQA}
for more references). Moreover, in the time-dependent case, with a vanishing
electromagnetic potential our Dirac operator is real quaternionic.

Here we exploit another attractive facet of our biquaternionic Dirac equation.
In the two-dimensional case it decouples into two separate Vekua equations. In
general these Vekua equations are bicomplex. However we show that the whole
theory of pseudoanalytic functions without modifications can be applied to
these equations under a certain not restrictive condition. As an example we
formulate the similarity principle which is the central reason why a
pseudoanalytic function and as a consequence a spinor field depending on two
space variables share many of the properties of analytic functions; e.g., they
are either identically zero or have isolated zeros. In this way more results
of the theory developed in \cite{Vekua} and in posterior works (see, e.g.,
\cite{Begehr} and \cite{Tutschke}) can be applied to the two-dimensional Dirac
equation with a scalar and an electromagnetic potentials. Nevertheless in the
present work we concentrate on another non-trivial and surprising consequence
of the established relation with pseudoanalytic functions. Consider the Dirac
equation with a scalar potential depending on one variable with fixed energy
and mass. In general this equation cannot be solved explicitly even if one
looks for wave functions of one variable. None the less the result of this
work is an algorithmically simple procedure for obtaining in explicit form a
complete system of exact solutions depending on two variables for such Dirac
equation. This system of solutions is a generalization of the system of powers
$1,z,z^{2},\ldots$ in complex analysis and as such they are not appropriate
for studying the Dirac equation on the whole plane. However the very fact that
it is always possible to obtain explicitly a complete system of exact
solutions of the Dirac equation with scalar potential of one variable as well
as the hope to be able to obtain explicitly not only the generalizations of
positive powers but also those of the negative ones makes in our opinion this
approach attractive and promising. The system of exact solutions for the Dirac
equation with a one-dimensional scalar potential is obtained due to the
proposed reduction of the Dirac equation to Vekua equations and due to L.
Bers' theory of Taylor series in formal powers.

In Section 2 we introduce notations. In Section 3 we give the biquaternionic
reformulation of the Dirac equation. Let us emphasize that our biquaternionic
Dirac equation is completely equivalent to the \textquotedblleft
traditional\textquotedblright\ Dirac equation written in $\gamma$-matrices, we
have a simple matrix transformation giving us a relation between their
solutions. In Section 4 we show that in a two-dimensional situation the Dirac
equation with a scalar and an electromagnetic potentials decouples into a pair
of bicomplex Vekua equations. We establish that if one of the coefficients in
such Vekua equation has not zeros and does not turn into a zero divisor at any
point of the domain of interest, the solutions will not be zero divisors
either, and the whole theory of generalized analytic functions without
modifications is applicable to the bicomplex Vekua equation.

In Section 5 we adapt some definitions and results from L. Bers' theory to
bicomplex pseudoanalytic functions. Section 6 is dedicated to a special class
of Vekua equations which have been studied recently (see \cite{KrBers} and
\cite{Krpseudoan}) due to their close relation to stationary Schr\"{o}dinger
equations. In Section 7 we show that the Dirac equation with a scalar
potential depending on one space variable can be represented as a Vekua
equation from the special class mentioned above. Here we should notice that
the case of the scalar potential is only an example. The same is true, for
example, for the electric potential. To the Vekua equation we apply L. Bers'
procedure for constructing corresponding formal powers which as was mentioned
above are exact solutions of the Vekua equation and generalize the system of
analytic functions $1,z,z^{2},\ldots$. With their aid any regular solution of
the Vekua equation can be represented by its Taylor series in formal powers.

\section{Preliminaries}

We denote by $\mathbb{H}(\mathbb{C})$ the algebra of complex quaternions (=
biquaternions). The elements of $\mathbb{H}(\mathbb{C})$ have the form
$q=\sum_{k=0}^{3}q_{k}e_{k}$ where $\left\{  q_{k}\right\}  \subset
\mathbb{C},$ $e_{0}$ is the unit and $\left\{  e_{k}|k=1,2,3\right\}  $ are
the standard quaternionic imaginary units.

We denote the imaginary unit in $\mathbb{C}$\ by $i$ as usual. By definition
$i$ commutes with $e_{k},$ $k=\overline{0,3}.$We will use also the vector
representation of $q\in\mathbb{H}(\mathbb{C}):$ $q=Sc(q)+Vec(q)$, where
$Sc(q)=q_{0}$ and $Vec(q)=\overrightarrow{q}=\sum_{k=1}^{3}q_{k}e_{k}.$ The
quaternionic conjugation is defined as follows $\overline{q}=q_{0}%
-\overrightarrow{q}$.

By $M^{p}$ we denote the operator of multiplication by $p$ from the right hand
side
\[
M^{p}q=q\cdot p.
\]
More information on complex quaternions the interested reader can find, e.g.,
in \cite{AQA} or \cite{KSbook}.

Let $q$ be a complex quaternion valued differentiable function of
$\mathbf{x}=(x_{1},x_{2},x_{3})$. Denote
\[
Dq=\sum_{k=1}^{3}e_{k}\frac{\partial}{\partial x_{k}}q.
\]
This operator is called sometimes the Moisil-Theodorescu operator or the Dirac
operator but the true is that it was introduced already by W.R. Hamilton
himself and studied in a great number of works (see, e.g., \cite{Dzhuraev},
\cite{GS1}, \cite{GS2}, \cite{AQA}, \cite{KSbook}).

\section{Quaternionic reformulation of the Dirac equation}

Consider the Dirac operator with scalar and electromagnetic potentials%
\[
\mathbb{D}=\gamma_{0}\partial_{t}+\sum_{k=1}^{3}\gamma_{k}\partial
_{k}+i\left(  m+p_{el}\gamma_{0}+\sum_{k=1}^{3}A_{k}\gamma_{k}+p_{sc}\right)
\]
where $\gamma_{j},$ $j=0,1,2,3$ are usual $\gamma$-matrices (see, e.g.,
\cite{BD}, \cite{Thaller}), $m\in\mathbb{R}$, $p_{el}$, $A_{k}$ and $p_{sc}$
are real valued functions.

In \cite{Krbag} a simple matrix transformation was obtained which allows us to
rewrite the classical Dirac equation in quaternionic terms.

Let us introduce an auxiliary notation $\widetilde{f}:=f(t,x_{1},x_{2}%
,-x_{3})$. The domain $\widetilde{G}$ is assumed to be obtained from the
domain $G\subset\mathbb{R}^{4}$ by the reflection $x_{3}\rightarrow-x_{3}$.
The transformation announced above we denote as $\mathcal{A}$ and define it in
the following way. A function $\Phi:G\subset\mathbb{R}^{4}\rightarrow
\mathbb{C}^{4}$ is transformed into a function $F:\widetilde{G}\subset
\mathbb{R}^{4}\rightarrow\mathbb{H(C)}$ by the rule
\[
F=\mathcal{A}[\Phi]:=\frac{1}{2}\left(  -(\widetilde{\Phi}_{1}-\widetilde
{\Phi}_{2})e_{0}+i(\widetilde{\Phi}_{0}-\widetilde{\Phi}_{3})e_{1}%
-(\widetilde{\Phi}_{0}+\widetilde{\Phi}_{3})e_{2}+i(\widetilde{\Phi}%
_{1}+\widetilde{\Phi}_{2})e_{3}\right)  .
\]
The inverse transformation $\mathcal{A}^{-1}$ is defined as follows
\[
\Phi=\mathcal{A}^{-1}[F]=(-i\widetilde{F}_{1}-\widetilde{F}_{2},-\widetilde
{F}_{0}-i\widetilde{F}_{3},\widetilde{F}_{0}-i\widetilde{F}_{3},i\widetilde
{F}_{1}-\widetilde{F}_{2})^{T}.
\]
Let us present the introduced transformations in a more explicit matrix form
which relates the components of a $\mathbb{C}^{4}$-valued function $\ \Phi$
with the components of an $\mathbb{H(C)}$-valued function $F$:
\[
F=\mathcal{A}[\Phi]=\frac{1}{2}\left(
\begin{array}
[c]{rrrr}%
0 & -1 & 1 & 0\\
i & 0 & 0 & -i\\
-1 & 0 & 0 & -1\\
0 & i & i & 0
\end{array}
\right)  \left(
\begin{array}
[c]{c}%
\widetilde{\Phi}_{0}\\
\widetilde{\Phi}_{1}\\
\widetilde{\Phi}_{2}\\
\widetilde{\Phi}_{3}%
\end{array}
\right)
\]
and%

\[
\Phi=\mathcal{A}^{-1}[{F}]=\left(
\begin{array}
[c]{rrrr}%
0 & -i & -1 & 0\\
-1 & 0 & 0 & -i\\
1 & 0 & 0 & -i\\
0 & i & -1 & 0
\end{array}
\right)  \left(
\begin{array}
[c]{c}%
\widetilde{F}_{0}\\
\widetilde{F}_{1}\\
\widetilde{F}_{2}\\
\widetilde{F}_{3}%
\end{array}
\right)  .
\]

\bigskip Denote
\[
R=D-\partial_{t}M^{e_{1}}+\mathbf{a}+M^{-i(\widetilde{p}_{el}e_{1}%
-i(\widetilde{p}_{sc}+m)e_{2})}%
\]
where $\mathbf{a}=i(\widetilde{A}_{1}e_{1}+\widetilde{A}_{2}e_{2}%
-\widetilde{A}_{3}e_{3})$. The following equality holds \cite{AQA}%
\[
R=\mathcal{A}\gamma_{1}\gamma_{2}\gamma_{3}\mathbb{D}\mathcal{A}^{-1}.
\]
That is, a $\mathbb{C}^{4}$-valued function $\Phi$ is a solution of the
equation
\[
\mathbb{D}\Phi=0\qquad\text{in }G
\]
iff the complex quaternionic function $F=\mathcal{A}\Phi$ is a solution of the
quaternionic equation
\[
RF=0\qquad\text{in }\widetilde{G}.
\]
Note that in the absence of the electromagnetic potential the operator $R$
becomes real quaternionic which is an important property (see \cite{KrRam}).

In what follows we assume that potentials are time-independent and consider
solutions with fixed energy: $\Phi(t,\mathbf{x})=\Phi_{\omega}(\mathbf{x}%
)e^{i\omega t}$. The equation for $\Phi_{\omega}$ has the form%
\begin{equation}
\mathbb{D}_{\omega}\Phi_{\omega}=0\qquad\text{in }\widehat{G}
\label{DiracOmega}%
\end{equation}
where $\widehat{G}$ is a domain in $\mathbb{R}^{3}$,%
\[
\mathbb{D}_{\omega}=i\omega\gamma_{0}+\sum_{k=1}^{3}\gamma_{k}\partial
_{k}+i\left(  m+p_{el}\gamma_{0}+\sum_{k=1}^{3}A_{k}\gamma_{k}+p_{sc}\right)
.
\]
We have
\[
R_{\omega}=\mathcal{A}\gamma_{1}\gamma_{2}\gamma_{3}\mathbb{D}_{\omega
}\mathcal{A}^{-1},
\]
where
\[
R_{\omega}=D+\mathbf{a}+M^{\mathbf{b}}%
\]
with $\mathbf{b}=-i((\widetilde{p}_{el}+\omega)e_{1}-i(\widetilde{p}%
_{sc}+m)e_{2})$. Thus, equation (\ref{DiracOmega}) turns into the complex
quaternionic equation
\begin{equation}
R_{\omega}q=0 \label{ROmega}%
\end{equation}
where $q$ is a complex quaternion valued function.

\section{The Dirac equation in a two-dimensional case as a bicomplex Vekua
equation\label{SectDiracVekua}}

Let us introduce the following notation. For any complex quaternion $q$ we
denote by $Q_{1}$ and $Q_{2}$ its bicomplex components:%
\[
Q_{1}=q_{0}+q_{3}e_{3}\qquad\text{and\qquad}Q_{2}=q_{2}-q_{1}e_{3}.
\]
Then $q$ can be represented as follows $q=Q_{1}+Q_{2}e_{2}$. For the operator
$D$ we have $D=D_{1}+D_{2}e_{2}$ with $D_{1}=e_{3}\partial_{3}$ and
$D_{2}=\partial_{2}-\partial_{1}e_{3}$. Notice that $\mathbf{b}=Be_{2}$ with
$B=-(\widetilde{p}_{sc}+m)+i(\widetilde{p}_{el}+\omega)e_{3}$, $\mathbf{a}%
=A_{1}+A_{2}e_{2}$ with $A_{1}=a_{3}e_{3}$ and $A_{2}=a_{2}-a_{1}e_{3}$.

We obtain that equation (\ref{ROmega}) is equivalent to the system%
\begin{equation}
D_{1}Q_{1}-D_{2}\overline{Q}_{2}+A_{1}Q_{1}-A_{2}\overline{Q}_{2}-\overline
{B}Q_{2}=0, \label{Diracsys1}%
\end{equation}%
\begin{equation}
D_{2}\overline{Q}_{1}+D_{1}Q_{2}+A_{2}\overline{Q}_{1}+A_{1}Q_{2}+BQ_{1}=0,
\label{Diracsys2}%
\end{equation}
where $Q_{1}$ and $Q_{2}$ are bicomplex components of $q$. We stress \ that
the system (\ref{Diracsys1}), (\ref{Diracsys2}) is equivalent to the Dirac
equation in $\gamma$-matrices (\ref{DiracOmega}).

Let us suppose all fields in our model to be independent of $x_{3}$, and
$A_{1}=a_{3}e_{3}\equiv0$. Then the system (\ref{Diracsys1}), (\ref{Diracsys2}%
) decouples, and we obtain two separate bicomplex equations%
\[
\overline{D}_{2}Q_{2}=-\overline{A}_{2}Q_{2}-B\overline{Q}_{2}%
\]
and%
\[
\overline{D}_{2}Q_{1}=-\overline{A}_{2}Q_{1}-\overline{B}\overline{Q}_{1}.
\]
Denote $\overline{\partial}=\overline{D}_{2}$, $a=-\overline{A}_{2}$, $b=-B$,
$w=Q_{2}$, $W=Q_{1}$, $z=x+y\mathbf{k}$, where $x=x_{2}$, $y=x_{1}$ and for
convenience we denote $\mathbf{k}=e_{3}$. Then we reduce the Dirac equation
with electromagnetic and scalar potentials independent of $x_{3}$ to a pair of
Vekua-type equations%
\begin{equation}
\overline{\partial}w=aw+b\overline{w} \label{Vekua1}%
\end{equation}
and
\begin{equation}
\overline{\partial}W=aW+\overline{bW}. \label{Vekua2}%
\end{equation}
The difference between the bicomplex equations (\ref{Vekua1}), (\ref{Vekua2})
\ and the usual complex Vekua equations is revealed if only $w$ or $W$ can
take values equal to bicomplex zero divisors (otherwise equations
(\ref{Vekua1}), (\ref{Vekua2}) can be analyzed following Bers-Vekua theory
\cite{Berskniga}, \cite{Vekua}). Let us study this possibility with the aid of
the following pair of projection operators%
\[
P^{+}=\frac{1}{2}(1+i\mathbf{k})\qquad\text{and\qquad}P^{-}=\frac{1}%
{2}(1-i\mathbf{k}).
\]
The set of bicomplex zero divisors, that is of nonzero elements $q=q_{0}%
+q_{1}\mathbf{k}$, $\left\{  q_{0},q_{1}\right\}  \subset\mathbb{C}$ such
that
\begin{equation}
q\overline{q}=\left(  q_{0}+q_{1}\mathbf{k}\right)  \left(  q_{0}%
-q_{1}\mathbf{k}\right)  =0 \label{Defzerodiv}%
\end{equation}
we denote by $\mathfrak{S}$.

\begin{lemma}
Let $q$ be a bicomplex number of the form $q=q_{0}+q_{1}\mathbf{k}$, $\left\{
q_{0},q_{1}\right\}  \subset\mathbb{C}$. If $q\in\mathfrak{S}$ then
$q=2P^{+}q_{0}$ or $q=2P^{-}q_{0}$.
\end{lemma}

\begin{proof}
From (\ref{Defzerodiv}) it follows that $q_{0}^{2}+q_{1}^{2}=0$ which gives us
that $q_{1}=\pm iq_{0}$. That is $q=q_{0}(1+i\mathbf{k})$ or $q=q_{0}%
(1-i\mathbf{k})$.
\end{proof}

For other results on bicomplex numbers we refer to \cite{RochonTrembl}.

Let $\Omega$ denote a bounded, simply connected domain in the plane of the
variable $z$.

\begin{theorem}
Let $b(z)\notin\mathfrak{S}\cup\left\{  0\right\}  $, $\forall z\in\Omega$ and
$w$, $W$ be solutions of (\ref{Vekua1}) and (\ref{Vekua2}) respectively. Then
$w(z)\notin\mathfrak{S}$ and $W(z)\notin\mathfrak{S}$, $\forall z\in\Omega$.
\end{theorem}

\begin{proof}
Assume that $w(z)\in\mathfrak{S}$ for some $z\in\Omega$. For definiteness let
$w(z)=2P^{+}w_{0}(z)$. Then from (\ref{Vekua1}) we have
\[
\overline{\partial}P^{+}w_{0}=aP^{+}w_{0}+bP^{-}w_{0}.
\]
Applying $P^{-}$ to this equality we find that $P^{-}b=0$ which is a contradiction.
\end{proof}

Thus, if the coefficient $b$ does not have zeros and does not turn into a zero
divisor at any point of the domain of interest, the solutions of
(\ref{Vekua1}) and (\ref{Vekua2}) will not be zero divisors either and the
whole theory of pseudoanalytic functions is applicable without changes to the
bicomplex equations (\ref{Vekua1}) and (\ref{Vekua2}). As an example let us
formulate one of the main results of the theory, the similarity principle
which is the basic tool for studying the distribution of zeros and of
singularities of pseudoanalytic functions as well as boundary value problems
\cite{Vekua}.

\begin{theorem}
Let $w$ be a regular solution of (\ref{Vekua1}) in a domain $\Omega$ and let
$b(z)\notin\mathfrak{S}\cup\left\{  0\right\}  $, $\forall z\in\Omega$. Then
the bicomplex function $\Phi=w\cdot e^{h}$, where
\[
h(z)=\frac{1}{2\pi}\int_{\Omega}\frac{g(\tau)d\tau}{\tau-z},
\]%
\[
g(z)=\left\{
\begin{array}
[c]{l}%
a(z)+b(z)\frac{\overline{w}(z)}{w(z)}\text{\quad if }w(z)\neq0,\quad
z\in\Omega,\\
a(z)+b(z)\text{\quad if }w(z)=0,\quad z\in\Omega
\end{array}
\right.
\]
is a solution of the equation $\overline{\partial}\Phi=0$ in $\Omega$.
\end{theorem}

The proof of this theorem is completely analogous to that given in
\cite{Vekua}. It would be interesting to extend this result to the case of $b$
being a zero divisor in the whole domain $\Omega$ or in some points.

This theorem opens the way to generalize many classical results from theory of
analytic functions to the case of solutions of equations (\ref{Vekua1}) and
(\ref{Vekua2}) by analogy with \cite{Vekua}. Nevertheless in the present work
we prefer to explore another possibility. Namely, we show how the application
of Bers' theory of pseudoanalytic functions allows us to obtain explicitly a
complete system of solutions of the Dirac equation with a scalar potential
depending on one variable.

\section{Some definitions and results from Bers' theory for bicomplex
pseudoanalytic functions}

\subsection{Generating pair, derivative and antiderivative}

Following \cite{Berskniga} we introduce the notion of a bicomplex generating pair.

\begin{definition}
A pair of bicomplex functions $F=F_{0}+F_{1}\mathbf{k}$ and $G=G_{0}%
+G_{1}\mathbf{k}$, possessing in $\Omega$ partial derivatives with respect to
the real variables $x$ and $y$ is said to be a generating pair if it satisfies
the inequality%
\[
\operatorname{Vec}(\overline{F}G)\neq0\qquad\text{in }\Omega.
\]
The following expressions are called characteristic coefficients of the pair
$(F,G)$
\[
a_{(F,G)}=-\frac{\overline{F}G_{\overline{z}}-F_{\overline{z}}\overline{G}%
}{F\overline{G}-\overline{F}G},\qquad b_{(F,G)}=\frac{FG_{\overline{z}%
}-F_{\overline{z}}G}{F\overline{G}-\overline{F}G},
\]

\end{definition}

\[
A_{(F,G)}=-\frac{\overline{F}G_{z}-F_{z}\overline{G}}{F\overline{G}%
-\overline{F}G},\qquad B_{(F,G)}=\frac{FG_{z}-F_{z}G}{F\overline{G}%
-\overline{F}G},
\]
where the subindex $\overline{z}$ or $z$ means the application of
$\overline{\partial}$ or $\partial$ respectively.

Every bicomplex function $W$ defined in a subdomain of $\Omega$ admits the
unique representation $W=\phi F+\psi G$ where the functions $\phi$ and $\psi$
are complex valued.

The $(F,G)$-derivative $\overset{\cdot}{W}=\frac{d_{(F,G)}W}{dz}$ of a
function $W$ exists and has the form
\begin{equation}
\overset{\cdot}{W}=\phi_{z}F+\psi_{z}G=W_{z}-A_{(F,G)}W-B_{(F,G)}\overline{W}
\label{FGder}%
\end{equation}
if and only if
\begin{equation}
\phi_{\overline{z}}F+\psi_{\overline{z}}G=0. \label{phiFpsiG}%
\end{equation}
This last equation can be rewritten in the following form%
\[
W_{\overline{z}}=a_{(F,G)}W+b_{(F,G)}\overline{W}%
\]
which we call the bicomplex Vekua equation. Solutions of this equation are
called $(F,G)$-pseudoanalytic functions.

\begin{remark}
The functions $F$ and $G$ are $(F,G)$-pseudoanalytic, and $\overset{\cdot}%
{F}\equiv\overset{\cdot}{G}\equiv0$.
\end{remark}

\begin{definition}
\label{DefSuccessor}Let $(F,G)$ and $(F_{1},G_{1})$ - be two generating pairs
in $\Omega$. $(F_{1},G_{1})$ is called \ successor of $(F,G)$ and $(F,G)$ is
called predecessor of $(F_{1},G_{1})$ if%
\[
a_{(F_{1},G_{1})}=a_{(F,G)}\qquad\text{and}\qquad b_{(F_{1},G_{1})}%
=-B_{(F,G)}\text{.}%
\]

\end{definition}

The importance of this definition becomes obvious from the following statement.

\begin{theorem}
\label{ThBersDer}Let $W$ be an $(F,G)$-pseudoanalytic function and let
$(F_{1},G_{1})$ be a successor of $(F,G)$. Then $\overset{\cdot}{W}$ is an
$(F_{1},G_{1})$-pseudoanalytic function.
\end{theorem}

\begin{definition}
\label{DefAdjoint}Let $(F,G)$ be a generating pair. Its adjoint generating
pair $(F,G)^{\ast}=(F^{\ast},G^{\ast})$ is defined by the formulas%
\[
F^{\ast}=-\frac{2\overline{F}}{F\overline{G}-\overline{F}G},\qquad G^{\ast
}=\frac{2\overline{G}}{F\overline{G}-\overline{F}G}.
\]

\end{definition}

The $(F,G)$-integral is defined as follows
\[
\int_{\Gamma}Wd_{(F,G)}z=\frac{1}{2}\left(  F(z_{1})\operatorname{Sc}%
\int_{\Gamma}G^{\ast}Wdz+G(z_{1})\operatorname{Sc}\int_{\Gamma}F^{\ast
}Wdz\right)
\]
where $\Gamma$ is a rectifiable curve leading from $z_{0}$ to $z_{1}$.

If $W=\phi F+\psi G$ is an $(F,G)$-pseudoanalytic function where $\phi$ and
$\psi$ are complex valued functions then
\begin{equation}
\int_{z_{0}}^{z}\overset{\cdot}{W}d_{(F,G)}z=W(z)-\phi(z_{0})F(z)-\psi
(z_{0})G(z), \label{FGAnt}%
\end{equation}
and as $\overset{\cdot}{F}=\overset{}{\overset{\cdot}{G}=}0$, this integral is
path-independent and represents the $(F,G)$-antiderivative of $\overset{\cdot
}{W}$.

\subsection{Generating sequences and Taylor series in formal
powers\label{SubsectGenSeq}}

\begin{definition}
\label{DefSeq}A sequence of generating pairs $\left\{  (F_{m},G_{m})\right\}
$, $m=0,\pm1,\pm2,\ldots$ , is called a generating sequence if $(F_{m+1}%
,G_{m+1})$ is a successor of $(F_{m},G_{m})$. If $(F_{0},G_{0})=(F,G)$, we say
that $(F,G)$ is embedded in $\left\{  (F_{m},G_{m})\right\}  $.
\end{definition}

\begin{theorem}
Let \ $(F,G)$ be a generating pair in $\Omega$. Let $\Omega_{1}$ be a bounded
domain, $\overline{\Omega}_{1}\subset\Omega$. Then $(F,G)$ can be embedded in
a generating sequence in $\Omega_{1}$.
\end{theorem}

\begin{definition}
A generating sequence $\left\{  (F_{m},G_{m})\right\}  $ is said to have
period $\mu>0$ if $(F_{m+\mu},G_{m+\mu})$ is equivalent to $(F_{m},G_{m})$
that is their characteristic coefficients coincide.
\end{definition}

Let $W$ be an $(F,G)$-pseudoanalytic function. Using a generating sequence in
which $(F,G)$ is embedded we can define the higher derivatives of $W$ by the
recursion formula%
\[
W^{[0]}=W;\qquad W^{[m+1]}=\frac{d_{(F_{m},G_{m})}W^{[m]}}{dz},\quad
m=1,2,\ldots\text{.}%
\]

\begin{definition}
\label{DefFormalPower}The formal power $Z_{m}^{(0)}(a,z_{0};z)$ with center at
$z_{0}\in\Omega$, coefficient $a$ and exponent $0$ is defined as the linear
combination of the generators $F_{m}$, $G_{m}$ with complex constant
coefficients $\lambda$, $\mu$ chosen so that $\lambda F_{m}(z_{0})+\mu
G_{m}(z_{0})=a$. The formal powers with exponents $n=1,2,\ldots$ are defined
by the recursion formula%
\begin{equation}
Z_{m}^{(n+1)}(a,z_{0};z)=(n+1)\int_{z_{0}}^{z}Z_{m+1}^{(n)}(a,z_{0}%
;\zeta)d_{(F_{m},G_{m})}\zeta. \label{recformula}%
\end{equation}

\end{definition}

This definition implies the following properties.

\begin{enumerate}
\item $Z_{m}^{(n)}(a,z_{0};z)$ is an $(F_{m},G_{m})$-pseudoanalytic function
of $z$.

\item If $a^{\prime}$ and $a^{\prime\prime}$ are complex constants, then
\[
Z_{m}^{(n)}(a^{\prime}+\mathbf{k}a^{\prime\prime},z_{0};z)=a^{\prime}%
Z_{m}^{(n)}(1,z_{0};z)+a^{\prime\prime}Z_{m}^{(n)}(\mathbf{k},z_{0};z).
\]

\item The formal powers satisfy the differential relations%
\[
\frac{d_{(F_{m},G_{m})}Z_{m}^{(n)}(a,z_{0};z)}{dz}=nZ_{m+1}^{(n-1)}%
(a,z_{0};z).
\]

\item The asymptotic formulas
\[
Z_{m}^{(n)}(a,z_{0};z)\sim a(z-z_{0})^{n},\quad z\rightarrow z_{0}%
\]
hold.
\end{enumerate}

Assume now that
\begin{equation}
W(z)=\sum_{n=0}^{\infty}Z^{(n)}(a,z_{0};z) \label{series}%
\end{equation}
where the absence of the subindex $m$ means that all the formal powers
correspond to the same generating pair $(F,G),$ and the series converges
uniformly in some neighborhood of $z_{0}$. It can be shown that the uniform
limit of pseudoanalytic functions is pseudoanalytic, and that a uniformly
convergent series of $(F,G)$-pseudoanalytic functions can be $(F,G)$%
-differentiated term by term. Hence the function $W$ in (\ref{series}) is
$(F,G)$-pseudoanalytic and its $r$th derivative admits the expansion
\[
W^{[r]}(z)=\sum_{n=r}^{\infty}n(n-1)\cdots(n-r+1)Z_{r}^{(n-r)}(a_{n}%
,z_{0};z).
\]
From this the Taylor formulas for the coefficients are obtained%
\begin{equation}
a_{n}=\frac{W^{[n]}(z_{0})}{n!}. \label{Taylorcoef}%
\end{equation}

\begin{definition}
Let $W(z)$ be a given $(F,G)$-pseudoanalytic function defined for small values
of $\left\vert z-z_{0}\right\vert $. The series%
\begin{equation}
\sum_{n=0}^{\infty}Z^{(n)}(a,z_{0};z) \label{Taylorseries}%
\end{equation}
with the coefficients given by (\ref{Taylorcoef}) is called the Taylor series
of $W$ at $z_{0}$, formed with formal powers.
\end{definition}

The Taylor series always represents the function asymptotically:%
\begin{equation}
W(z)-\sum_{n=0}^{N}Z^{(n)}(a,z_{0};z)=O\left(  \left\vert z-z_{0}\right\vert
^{N+1}\right)  ,\quad z\rightarrow z_{0}, \label{asympt}%
\end{equation}
for all $N$. This implies (since a pseudoanalytic function can not have a zero
of arbitrarily high order without vanishing identically) that the sequence of
derivatives $\left\{  W^{[n]}(z_{0})\right\}  $ determines the function $W$ uniquely.

If the series (\ref{Taylorseries}) converges uniformly in a neighborhood of
$z_{0}$, it converges to the function $W$.

\begin{theorem}
\label{ThConvPer} The formal Taylor expansion (\ref{Taylorseries}) of a
pseudoanalytic function in formal powers defined by a periodic generating
sequence converges in some neighborhood of the center.
\end{theorem}

\section{Special class of Vekua equations}

The following important class of Vekua equations was considered in
\cite{Krpseudoan}. Let $f_{0}$ be a complex valued (with respect to $i$),
twice differentiable nonvanishing function defined on $\Omega$. Consider the
equation%
\begin{equation}
\overline{\partial}W=\frac{\overline{\partial}f_{0}}{f_{0}}\overline{W}%
\qquad\text{in }\Omega. \label{Vekuamain}%
\end{equation}
Denote $\nu_{1}=\Delta f_{0}/f_{0}$.

\begin{theorem}
\label{ThConjugate} \cite{Krpseudoan} If $W=W_{1}+W_{2}\mathbf{k}$ is a
solution of (\ref{Vekuamain}) then $W_{1}=\operatorname{Sc}W$ is a solution of
the stationary Schr\"{o}dinger equation
\begin{equation}
-\Delta W_{1}+\nu_{1}W_{1}=0\qquad\text{in }\Omega\label{Schr1}%
\end{equation}
and $W_{2}=\operatorname{Vec}W$ is a solution of the associated
Schr\"{o}dinger equation
\begin{equation}
-\Delta W_{2}+\nu_{2}W_{2}=0\qquad\text{in }\Omega\label{Schr2}%
\end{equation}
where $\nu_{2}=2(\overline{\partial}f_{0}\cdot\partial f_{0})/f_{0}^{2}%
-\nu_{1}$.
\end{theorem}

Moreover, in \cite{Krpseudoan} a simple formula was obtained which allows us
for any given solution $W_{1}$ of (\ref{Schr1}) to construct such a solution
$W_{2}$ of (\ref{Schr2}) that $W=W_{1}+W_{2}\mathbf{k}$ will be a solution of
(\ref{Vekuamain}) generalizing in this way the well known procedure for
constructing conjugate harmonic functions in complex analysis.

\section{Dirac equation with a scalar potential}

Let us show that the Dirac equation with a scalar potential depending on one
real variable reduces to a bicomplex Vekua equation of the form
(\ref{Vekuamain}).

Let $p_{sc}=p(x)$ and $p_{el}\equiv0,$ $A_{k}\equiv0,$ $k=1,2,3$. Then
according to Section \ref{SectDiracVekua} the Dirac equation is equivalent to
the pair of bicomplex Vekua equations%
\begin{equation}
\overline{\partial}w=b\overline{w} \label{Vekuasc1}%
\end{equation}
and%
\begin{equation}
\overline{\partial}W=\overline{bW} \label{Vekuasc2}%
\end{equation}
with $b=p(x)+m-i\omega\mathbf{k}$.

Let $f_{0}=e^{P(x)+mx+i\omega y}$, where $P$ is an antiderivative of $p$. Then
we have
\[
\overline{b}=\overline{\partial}f_{0}/f_{0}.
\]
Note that due to theorem \ref{ThConjugate} if the bicomplex function $W$ is a
solution of (\ref{Vekuasc2}) then the complex function $W_{1}%
=\operatorname{Sc}W$ is a solution of the stationary Schr\"{o}dinger equation
(\ref{Schr1}) where
\begin{equation}
\nu_{1}(x)=p^{\prime}(x)+(p(x)+m)^{2}-\omega^{2}, \label{nu1}%
\end{equation}
and the function $W_{2}=\operatorname{Vec}W$ is a solution of equation
(\ref{Schr2}) where
\begin{equation}
\nu_{2}(x)=-p^{\prime}(x)+(p(x)+m)^{2}-\omega^{2}. \label{nu2}%
\end{equation}
Let us notice that both Schr\"{o}dinger equations (\ref{Schr1}) and
(\ref{Schr2}) in this case admit separation of variables. Nevertheless this
does not imply they can be solved explicitly. In general this is not the case.
However we will show how using our approach and Bers' theory for both of them
one can construct in explicit form a locally complete system of exact solutions.

Consider equation (\ref{Vekuasc2}). It is easy to see that the pair of
functions%
\begin{equation}
F=f_{0}\qquad\text{and}\qquad G=\frac{\mathbf{k}}{f_{0}} \label{F,G}%
\end{equation}
represents a generating pair for (\ref{Vekuasc2}). Note that $F=e^{\sigma}$
and $G=e^{-\sigma}\mathbf{k}$, where $\sigma=\alpha(x)+\beta(y)$ and
$\alpha(x)=P(x)+mx$, $\beta(y)=i\omega y$. For a generating pair of such
special kind it is easy to construct a successor \cite{Berskniga}. Let
$\tau=-\alpha(x)+\beta(y)$. Then the pair $F_{1}=e^{\tau}$ and $G_{1}%
=e^{-\tau}\mathbf{k}$ is a successor of $(F,G)$. Moreover, $(F,G)$ is a
successor of $(F_{1},G_{1})$. Thus, for $(F,G)$ we obtain a complete periodic
generating sequence of a period $2$ in explicit form (for explicitly
constructed, in general non-periodic generating sequences in a far more
general situation we refer to \cite{Krpseudoan}).

The fact that we have a generating sequence in explicit form implies that we
are able to construct the corresponding formal powers of any order explicitly
and therefore to obtain a locally complete system of exact solutions of the
Dirac equation with a scalar potential depending on one variable as well as of
the stationary Schr\"{o}dinger equations (\ref{Schr1}) and (\ref{Schr2}) with
potentials (\ref{nu1}) and (\ref{nu2}) respectively.

As a first step we construct the adjoint generating pair (see definition
\ref{DefAdjoint}):%
\[
F^{\ast}=-f_{0}\mathbf{k}\qquad\text{and}\qquad G^{\ast}=\frac{1}{f_{0}}.
\]
Next, we write down the expression for the $(F,G)$-integral:%
\[
\int_{\Gamma}Wd_{(F,G)}z=\frac{1}{2}\left(  f_{0}(z_{1})\operatorname{Sc}%
\int_{\Gamma}\frac{W(z)}{f_{0}(z)}dz-\frac{\mathbf{k}}{f_{0}(z_{1}%
)}\operatorname{Sc}\int_{\Gamma}f_{0}(z)W(z)\mathbf{k}dz\right)  .
\]
By definition, the formal power $Z^{(0)}(a,z_{0};z)$ for equation
(\ref{Vekuasc2}) has the form%
\[
Z^{(0)}(a,z_{0};z)=\lambda F(z)+\mu G(z),
\]
where the complex constants $\lambda$ and $\mu$ are chosen so that $\lambda
F(z_{0})+\mu G(z_{0})=a$. That is,%
\[
Z^{(0)}(a,z_{0};z)=\lambda e^{P(x)+mx+i\omega y}+\mu e^{-(P(x)+mx+i\omega
y)}\mathbf{k}.
\]
In order to obtain $Z^{(1)}(a,z_{0};z)$ we should take the $(F,G)$-integral of
$Z_{1}^{(0)}(a,z_{0};z)$, where
\[
Z_{1}^{(0)}(a,z_{0};z)=\lambda_{1}F_{1}(z)+\mu_{1}G_{1}(z),
\]
with $\lambda_{1}F_{1}(z_{0})+\mu_{1}G_{1}(z_{0})=a$. Thus,
\[
Z^{(1)}(a,z_{0};z)=\int_{z_{0}}^{z}(\lambda_{1}F_{1}(\zeta)+\mu_{1}G_{1}%
(\zeta))d_{(F,G)}\zeta
\]%
\begin{align*}
&  =\frac{1}{2}\{e^{P(x)+mx+i\omega y}\operatorname{Sc}\int_{z_{0}}%
^{z}e^{-P(x^{\prime})-mx^{\prime}-i\omega y^{\prime}}(\lambda_{1}%
e^{-P(x^{\prime})-mx^{\prime}+i\omega y^{\prime}}+\mu_{1}e^{P(x^{\prime
})+mx^{\prime}-i\omega y^{\prime}}\mathbf{k})d\zeta\\
&  -e^{-P(x)-mx-i\omega y}\mathbf{k}\operatorname{Sc}\int_{z_{0}}%
^{z}e^{P(x^{\prime})+mx^{\prime}+i\omega y^{\prime}}\mathbf{k}(\lambda
_{1}e^{-P(x^{\prime})-mx^{\prime}+i\omega y^{\prime}}+\mu_{1}e^{P(x^{\prime
})+mx^{\prime}-i\omega y^{\prime}}\mathbf{k})d\zeta\}
\end{align*}%
\begin{align*}
&  =\frac{1}{2}\{e^{P(x)+mx+i\omega y}\operatorname{Sc}\int_{z_{0}}%
^{z}(\lambda_{1}e^{-2(P(x^{\prime})+mx^{\prime})}+\mu_{1}e^{-2i\omega
y^{\prime}}\mathbf{k})d\zeta\\
&  -e^{-P(x)-mx-i\omega y}\mathbf{k}\operatorname{Sc}\int_{z_{0}}^{z}%
(\lambda_{1}e^{2i\omega y^{\prime}}\mathbf{k}-\mu_{1}e^{2(P(x^{\prime
})+mx^{\prime})})d\zeta\}
\end{align*}
where $\zeta=x^{\prime}+y^{\prime}\mathbf{k}$.

For $Z^{(2)}(a,z_{0};z)$ by definition \ref{DefFormalPower} we have%
\begin{equation}
Z^{(2)}(a,z_{0};z)=2\int_{z_{0}}^{z}Z_{1}^{(1)}(a,z_{0};\zeta)d_{(F,G)}\zeta,
\label{Z2}%
\end{equation}
where $Z_{1}^{(1)}(a,z_{0};\zeta)$ in its turn can be found from the equality
\begin{equation}
Z_{1}^{(1)}(a,z_{0};z)=\int_{z_{0}}^{z}Z_{2}^{(0)}(a,z_{0};\zeta
)d_{(F_{1},G_{1})}\zeta. \label{Z11}%
\end{equation}
We note that due to periodicity of the generating sequence containing the
generating pair (\ref{F,G}),%
\[
Z_{2}^{(0)}(a,z_{0};\zeta)=Z^{(0)}(a,z_{0};\zeta).
\]
The adjoint pair for $(F_{1},G_{1})$ necessary for the $(F_{1},G_{1}%
)$-integral in (\ref{Z11}) has the form%
\[
F_{1}^{\ast}=-e^{\tau}\mathbf{k}\qquad\text{and}\qquad G_{1}^{\ast}=e^{-\tau
}.
\]
Thus,%
\[
Z_{1}^{(1)}(a,z_{0};z)=
\]%
\begin{align*}
&  =\frac{1}{2}\{e^{-P(x)-mx+i\omega y}\operatorname{Sc}\int_{z_{0}}%
^{z}e^{P(x^{\prime})+mx^{\prime}-i\omega y^{\prime}}(\lambda e^{P(x^{\prime
})+mx^{\prime}+i\omega y^{\prime}}+\mu e^{-P(x^{\prime})-mx^{\prime}-i\omega
y^{\prime}}\mathbf{k})d\zeta\\
&  -e^{P(x)+mx-i\omega y}\mathbf{k}\operatorname{Sc}\int_{z_{0}}%
^{z}e^{-P(x^{\prime})-mx^{\prime}+i\omega y^{\prime}}\mathbf{k}(\lambda
e^{P(x^{\prime})+mx^{\prime}+i\omega y^{\prime}}+\mu e^{-P(x^{\prime
})-mx^{\prime}-i\omega y^{\prime}}\mathbf{k})d\zeta\}
\end{align*}%
\begin{align*}
&  =\frac{1}{2}\{e^{-P(x)-mx+i\omega y}\operatorname{Sc}\int_{z_{0}}%
^{z}(\lambda e^{2(P(x^{\prime})+mx^{\prime})}+\mu e^{-2i\omega y^{\prime}%
}\mathbf{k})d\zeta\\
&  -e^{P(x)+mx-i\omega y}\mathbf{k}\operatorname{Sc}\int_{z_{0}}^{z}(\lambda
e^{2i\omega y^{\prime}}\mathbf{k}-\mu e^{-2(P(x^{\prime})+mx^{\prime})}%
)d\zeta\}.
\end{align*}
Substitution of this expression into (\ref{Z2}) gives us the formal power
$Z^{(2)}(a,z_{0};z)$, and this algorithmically simple procedure can be
continued indefinitely. As a result we obtain an infinite system of formal
powers which at least locally gives us a complete system of solutions of
(\ref{Vekuasc2}) in the sense that any regular solution of (\ref{Vekuasc2})
can be approximated arbitrarily closely by a finite linear combination of
formal powers (formula (\ref{asympt})). Moreover, as the corresponding
generating sequence is periodic, theorem \ref{ThConvPer} is valid, and
therefore we can guarantee the convergence of a Taylor expansion in the formal
powers to a corresponding solution of (\ref{Vekuasc2}) in some neighborhood of
$z_{0}$.

A similar procedure works also for equation (\ref{Vekuasc1}). Note that the
pair of functions $F_{1}\mathbf{k}=e^{\tau}\mathbf{k}$ and $G_{1}%
\mathbf{k}=-e^{-\tau}$ is a generating pair corresponding to (\ref{Vekuasc1}).

As any solution of the Schr\"{o}dinger equation (\ref{Schr1}) with the
potential $\nu_{1}$ defined by (\ref{nu1}) is the scalar part of some solution
of (\ref{Vekuasc2}), and any solution of (\ref{Schr2}) with the potential
(\ref{nu2}) is the vector part of some solution of (\ref{Vekuasc2}), the
scalar and the vector parts of the constructed system of formal powers give us
locally complete systems of solutions of (\ref{Schr1}) and (\ref{Schr2}) respectively.

This last result can also be interpreted in the following way. Consider the
equation
\begin{equation}
-\Delta f+\nu f=\omega^{2}f\qquad\text{in }\Omega\label{Schr}%
\end{equation}
where $f$ is a complex twice continuously differentiable function of two real
variables $x$ and $y$, and $\nu$ is a complex valued function of one real
variable $x$, $\omega$ is a complex constant. Suppose we are given a
particular solution $f_{0}=f_{0}(x)$ of the ordinary differential equation
\begin{equation}
-\frac{d^{2}f_{0}}{dx^{2}}+\nu f_{0}=0.\label{Schrord}%
\end{equation}
This implies that we are able to represent $\nu$ in the form $\nu=p^{\prime
}+p^{2}$ where $p=f_{0}^{\prime}/f_{0}$. Then we observe that (\ref{Schr}) is
precisely equation (\ref{Schr1}) with $m=0$ in (\ref{nu1}). Thus our result
means that if we are able to solve the ordinary differential equation
(\ref{Schrord}) then we can construct explicitly a locally complete system of
exact solutions to (\ref{Schr}) for any $\omega$. For this one should consider
the bicomplex Vekua equation (\ref{Vekuasc2}) and follow the procedure
described above for constructing the corresponding system of formal powers.
Then the scalar part of the system gives us a locally complete system of exact
solutions to (\ref{Schr}).

\end{document}